\documentclass[twocolumn,superscriptaddress,floatfix, nofootinbib]{revtex4-1}

\usepackage{etex}
\usepackage{ifpdf}
\usepackage{hyperref}
\usepackage{dcolumn}
\usepackage{url}
\usepackage{amsmath}
\usepackage{amscd}
\usepackage{amsfonts}
\usepackage{amssymb}
\usepackage{amsthm}
\usepackage{xfrac}
\usepackage{braket}
\usepackage{bm}   
\usepackage{verbatim}
\usepackage{stmaryrd}
\usepackage{xcolor}
\usepackage{setspace}
\usepackage{tikz}
\usetikzlibrary{arrows}
\usetikzlibrary{automata}
\usetikzlibrary{positioning}
\usetikzlibrary{plotmarks}
\usepackage{pgfplots}
\pgfplotsset{compat=newest}
\usepackage{floatrow}

\theoremstyle{plain}    
\theoremstyle{plain}    
\theoremstyle{plain}    
\theoremstyle{plain}    
\theoremstyle{plain}    
\theoremstyle{plain}    
\theoremstyle{plain}    
\theoremstyle{plain}    
\theoremstyle{plain}    
\theoremstyle{plain}    
\theoremstyle{plain}    
\theoremstyle{plain}    
\theoremstyle{plain}    

\colorlet {R_color}    {blue}
\colorlet {k_color}    {black!30!green}

\def\clap#1{\hbox to 0pt{\hss#1\hss}}

\newcommand{\Prob}      {\Pr} 

\begin{document}
\title{The evolution of lossy compression}
\author{Sarah E.\ Marzen}
\affiliation{Department of Physics, University of California at Berkeley, Berkeley, CA 94720}
\author{Simon DeDeo}\thanks{To whom correspondence should be addressed. {\tt sdedeo@indiana.edu}}
\affiliation{Cognitive Science Program \& Department of Informatics, Indiana University, 901 E 10th St, Bloomington, IN 47408}
\affiliation{Santa Fe Institute, 1399 Hyde Park Rd, Santa Fe, NM 87501}
\bibliographystyle{unsrt}

\begin{abstract}
\noindent
In complex environments, there are costs to both ignorance and perception. An organism needs to track fitness-relevant information about its world, but the more information it tracks, the more resources it must devote to memory and processing. Rate-distortion theory shows that, when errors are allowed, remarkably efficient internal representations can be found by biologically-plausible hill-climbing mechanisms. We identify two regimes: a high-fidelity regime where perceptual costs scale logarithmically with environmental complexity, and a low-fidelity regime where perceptual costs are, remarkably, independent of the environment. When environmental complexity is rising, Darwinian evolution should drive organisms to the threshold between the high- and low-fidelity regimes. Organisms that code efficiently will find themselves able to make, just barely, the most subtle distinctions in their environment. 
\end{abstract}
\keywords{lossy compression | evolution | rate-distortion | information theory | perception | signaling | neuroscience}

\maketitle

To survive, organisms must extract useful information from the environment. This is true over an individual's lifetime, when neural spikes~\cite{rieke1999spikes}, signaling molecules~\cite{tlusty2008rate, hancock2010cell}, or epigenetic markers~\cite{prohaska2010innovation} encode transient features, as well as at the population level and over generational timescales, where the genome can be understood as hard-wiring facts about the environments under which it evolved~\cite{krakauer2011darwinian}. Processing infrastructure may be built dynamically in response to environmental complexity~\cite{CNE:CNE902890115,Leggio200578,Simpson2011246,Scholz2015190}, but organisms cannot retain all potentially useful information because the real world is too complicated. Instead, they can reduce resource demands by tracking a smaller number of features~\cite{Crut92c, Bial00a, Gers13a,wolpert2014framework,sims2003implications}. 

When they do this, evolved organisms are expected to structure their perceptual systems to avoid dangerous confusions (not mistaking tigers for bushes) while strategically containing processing costs by allowing for ambiguity (using a single representation for both tigers and lions)---a form of {\it lossy} compression that avoids storing unnecessary and less-useful information. 

We use the informal language of mammalian perception for our examples here, but similar concerns apply to, for example, a cellular signaling system which might need to distinguish temperature signals from signs of low pH, while tolerating confusion of high temperature with low oxygenation. We include memory of both low-level percepts and the higher-level concepts they create and that play a role in decision-making~\cite{newsome1989neuronal,Ernst2004162,Silvanto2012840}. Memory costs include error-correction and circuit redundancy necessary to process and transmit in noisy systems~\cite{Shadlen1994569}. 

In order to quantify this tradeoff, we must first characterize the costs of confusion. We do so using a distortion measure, $d(x, \tilde{x})$, that represents the cost to the organism of mistaking one environmental state, $x$, for a different state, $\tilde{x}$.  When $d(x,\tilde{x})$ is large, mistaking state $x$ for state $\tilde{x}$ is costly.  This distortion measure need not be symmetric (it is more costly to mistake a tiger for a bush than to mistake a bush for a tiger) and not all off-diagonal elements need be large (it is not necessarily more costly to mistake a tiger for a lion).  We assume that $d(x,x)$ is zero for every state $x$---that with a well-chosen action policy, we can do no better perceptually than to faithfully record our environment. 


A great deal of effort has gone into choosing good distortion measures \cite{Tish00a, sims2003implications, wang2004image, Stil07c}.  It turns out that we can make surprisingly specific predictions about how the organism's memory scales with environmental complexity under mild assumptions about the substrate for memory storage, the distortion function and (implicitly) the organism's action policy, and the structure of the environment's pasts.

For any environment, there is a the minimal cost to misperceiving an environmental signal, $d_\mathrm{min}$, equal to $\min_{x\neq \tilde{x}} d(x,\tilde{x})$; pairs $x$ and $\tilde{x}$ that satisfy this bound are called ``minimal confounds''. When an organism attempts to achieve average distortion below this minimal level, we shall see that a critical transition occurs in how processing costs scale with complexity. This happens when $d_\mathrm{min}$ is independent of the size of the environment. The $d_\mathrm{min}$ threshold separates out a low- and high-fidelity regime. 

In a low-fidelity regime, when an organism's average distortion is larger than $d_\mathrm{min}$, increasing environmental complexity does not increase perceptual load. As the number of environmental states increases, innocuous synonyms accumulate. They do so sufficiently fast that an organism can continue to represent the fitness-relevant features within constant memory. 

It is only in a high-fidelity regime, when an organism attempts to achieve average distortions below $d_\mathrm{min}$, that memory load becomes sensitive to complexity. High-fidelity representations of the world do not scale; an organism that attempts to break this threshold will find that, when the number of environmental states increases, its own perceptual apparatus must also increase in size. As we shall see, the existence of this threshold has important implications for the evolution of perception.

\section{Rate Distortion Theory}

We quantify the tradeoff between memory storage and perceptual distortion using rate-distortion theory~\cite[Ch.~8]{Yeun08a}. Rate-distortion allows us to determine the extent to which the representations needed to completely describe a source (here, the organism's environment) can be compressed by selectively discarding information. 

This is lossy compression, in the sense that once the signal is encoded it becomes impossible to completely reconstruct the original. The process is represented by a codebook, $p(r|x)$, which describes the probability that an environmental state $x$ (drawn from the set $X$) is represented by a symbol $r$ (drawn from the set $\mathcal{R}$).

Careful choice of $p(r|x)$ allows for the organism to conserve resources that would otherwise be committed to the retention of information. Information theory provides a mechanism-independent lower bound on these costs. In particular, the (effective) number of states stored by an organism's codebook, per input from the environment, is lower-bounded by $2^{I[\mathcal{R};X]}$. Here, mutual information quantifies the drop in uncertainty about the environment given the perceptual state,
\begin{equation}
I[\mathcal{R};X] = H[X] - H[X|\mathcal{R}],
\end{equation}
where $H[X]$ is the entropy (uncertainty) of the environment, and $H[X|\mathcal{R}]$ is the conditional entropy, given knowledge of the perceptual state $r$. $I[\mathcal{R};X]$ quantifies the cost of memory for this lossy representation. It provides a strict lower bound to the costs of retention and storage, independent of mechanism. 

While biological systems are unlikely to precisely saturate this bound, evidence from a variety of sources shows that the pressure to store information as efficiently as possible, given fitness constraints, strongly influences evolved design, including the genetic code for amino acids~\cite{wagner2013robustness} and neural representations of visual scenes~\cite{olshausen1997sparse}. Simple arguments show that, for neural codes, $I[\mathcal{R};X]$ is an achievable lower bound on the number of neurons required to encode the environment at (an action-policy dependent) average distortion $D$ (see Supplementary Information).

To quantify over a variety of possible environments, we consider cases where the off-diagonal entries of $d(x,\tilde{x})$ are drawn from a distribution. Doing so allows us to study complicated environments with many degrees of freedom, and in which some mistakes are more costly than others. We begin with the case where $d(x,\tilde{x})$ entries are drawn independently; however, as we shall show, our bounds generalize beyond this and include environments with correlated penalties. 

Our problem is then that of finding an optimal codebook given an upper bound on average costs. We wish to build the most efficient perceptual system possible as long as, on average, the cost to the organism of errors induced by that codebook does not exceed some critical value, $D$. We can combine these two demands into a single minimization problem, where we find the minimal storage cost, $R(D)$, given the fitness constraint,
\begin{equation}
R(D) = \min_{p(r|x)~\textrm{~given~}~\mathbb{E}[d]\leq D} I[\mathcal{R};X],
\label{to_minimize}
\end{equation}
where $\mathbb{E}[d]$ is the average value of $d(x,r)$ given $p(x)$ (distribution over environmental states, taken to be uniform) and the codebook (to be found), $p(r|x)$.


The rate, $R(D)$, in bits, quantifies processing cost; $2^{R(D)}$ is the average cardinality of the organism's partition of the environment, which can be considered the effective number of internal representations, $N_\mathrm{eff}$, that the organism can use. The rate $R(D)$ is the best measure of processing cost when an organism's mechanisms include the ability to efficiently compress incoming sensory information. In our account of how perceptual costs scale with environmental complexity, we plot $R(D)$ as a function of the number of environmental states $N$. 

Eq.~\ref{to_minimize} is related to the information bottleneck method introduced by Ref.~\cite{Tish00a}. In that program, the distortion measure is itself an information theoretic quantity---a particularly useful choice for engineering and machine-learning applications~\cite{harremoes2007information}. Here, by contrast, we have introduced environmental structure; it is this external structure, in the form of $d(x,\tilde{x})$, that dictates the nature of the representations that the system uses.

Once we allow the average error rate, $D$, to become non-zero---\emph{i.e.}, once organisms can trade off perceptual and memory costs with accuracy---solutions of Eq.~\ref{to_minimize} provide a guide to how this tradeoff happens in practice. As we shall see, depending on accuracy demands, an individual will find itself in either a low- or high-fidelity regime, with qualitatively different properties.

\section{The Low-Fidelity Regime}


\begin{figure}
\begin{tabular}{cc}
\includegraphics[width=3in]{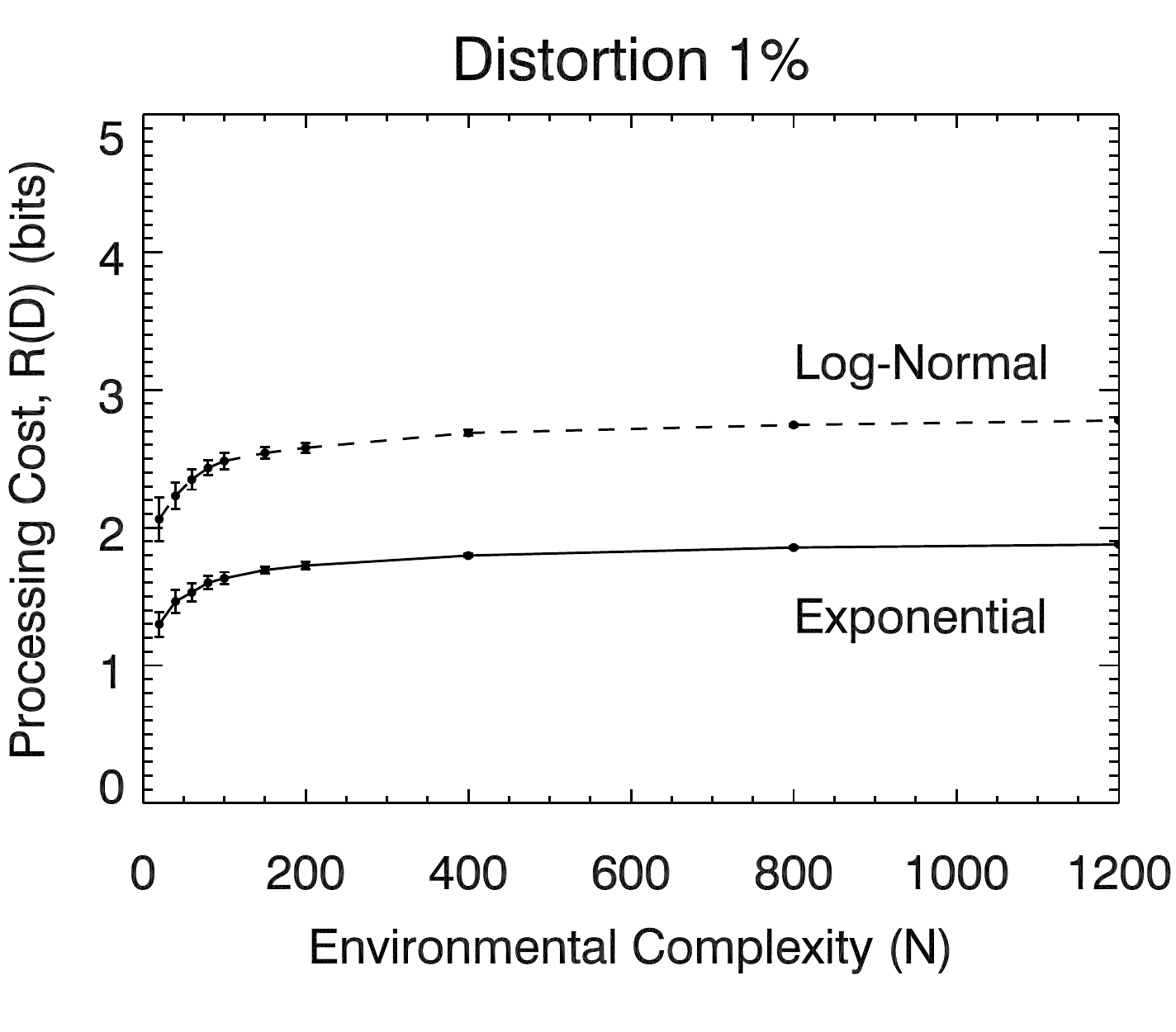} \\
\includegraphics[width=3in]{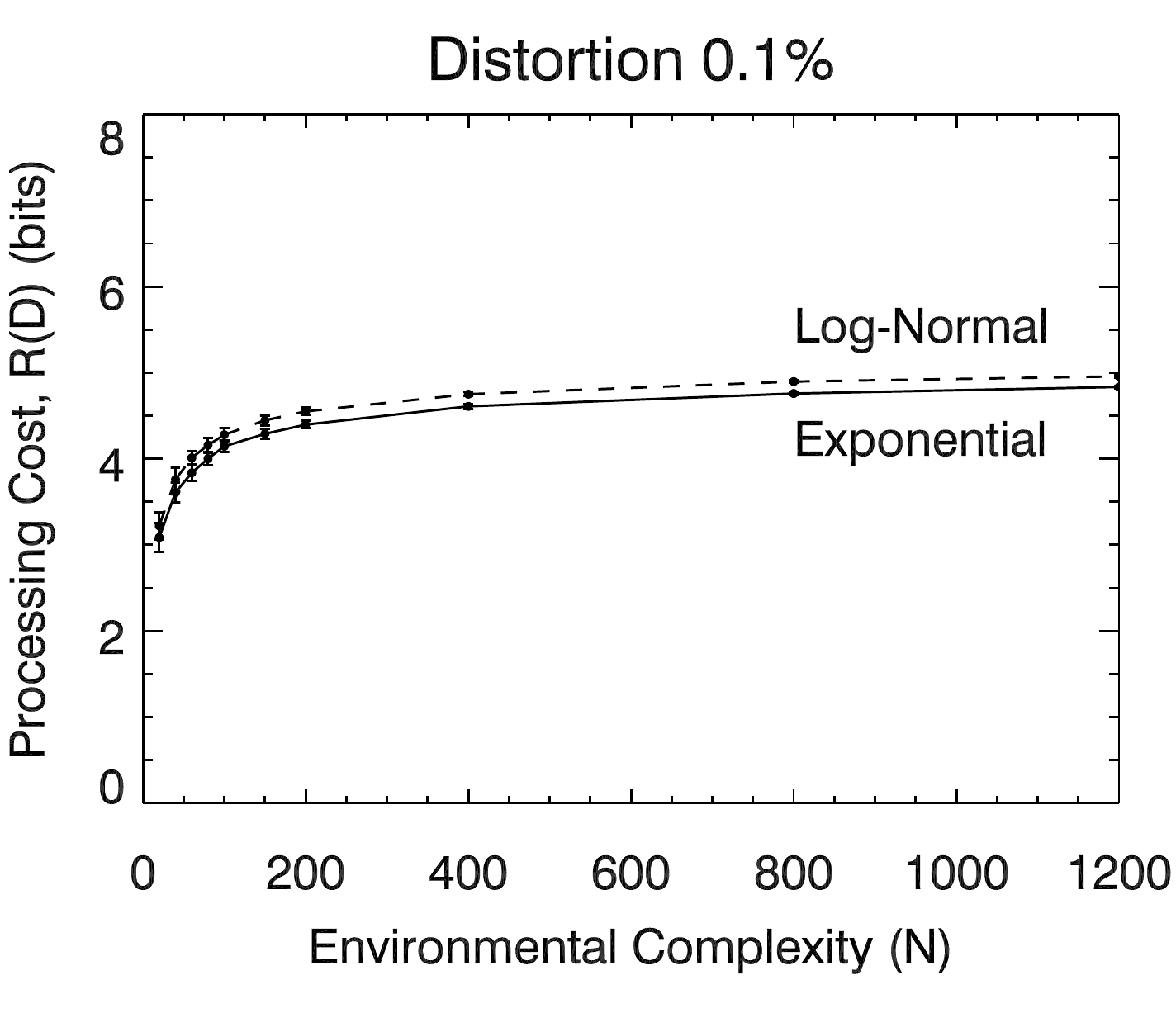}
\end{tabular}
\caption{\textbf{Fault-tolerant organisms can radically simplify their environment}. When an organism can tolerate errors in perception, large savings in storage are possible. Shown here is the scaling between environmental complexity, $N$, and processing costs $R(D)$, when we constrain the average distortion, $D$ to be $1\%$ and $0.1\%$ of the average cost of random guessing. We consider both environments with exponential and with log-normal (heavy-tailed) fitness; in both cases, $d_\mathrm{min}$ is equal to zero. A simple argument (see text) shows that in this low-fidelity regime, and as $N$ goes to infinity, perceptual demands asymptote to a constant. \label{exp_one}}
\end{figure}
Once accuracy costs can be greater than $d_\mathrm{min}$, we can achieve dramatically reduced bounds on storage. When organisms are tolerant to failure, they need only attend to a fraction of the environment. Furthermore, memory demands on a fault-tolerant organism increase far more slowly than the complexity of the environment. 

We study this first through simulations; Fig.~\ref{exp_one} shows results for optimal codebooks when error costs are exponentially and log-normally distributed; in both cases $d_\mathrm{min}$ is zero. An environment with hundreds of states can be compressed down to a few dozen at the cost of only 1\% in overall fitness.

Remarkably, when the organism's average error tolerance is greater than $d_\mathrm{min}$, the slow growth in processing costs, apparent in the simulations, asymptotes to a constant. The suggestive results of the simulations can be confirmed by simple analytic argument. For each of the $N$ environmental states, and overall error constraint $D$, there will be, on average, $NP(d< D)$ states that are allowable ambiguities---\emph{i.e.}, that the organism can take as synonyms for that percept. While this is not the most efficient coding, it provides an upper bound to the optimal rate $R(D)$.

The rate for such an encoding is approximately $\log{N}-\log{NP(d< D)}$, which becomes exact in the large $N$ limit. This means that both the encoding rate, $R(D)$ and the effective symbol size go to a constant; effective symbol size proportional to the inverse of the conditional probability distribution, $1/P(d< D)$, and $R(D)$ to its logarithm. More generally, $R(D)$ is bounded from above so that
\begin{equation}
R(D) \leq \frac{1}{N}\sum_x \log_2{\left(\frac{N}{N_D(x)}\right)},
\label{constant}
\end{equation}
where $N_D(x)$ is the number of synonyms for $x$ with distortion penalty less than $D$ (see Supplementary Information). This fact is independent of the underlying distribution. As long as $N_D$ is asymptotically proportional to $N$, this upper bound implies asymptotic insensitivity to the number of states of the environment. This is the case even when off diagonal entries are partially correlated, as long as correlation scales grow more slowly than environment size so that $N_D\propto N$.

For the exponential this upper bound means that ambiguities accumulate sufficiently fast that the organism will need at most 6.6 bits (for $D$ equal to 1\%) or 10 bits (for $D$ equal to $0.1\%$) even when the set of environmental states becomes arbitrarily large. 

This bound is not tight; as can be seen in Fig.~\ref{exp_one}, far better compressions are possible, and the system asymptotes at much lower values. When $N$ becomes large, these compressions are more reliably obtained (\emph{i.e.}, the variance in compression rate decreases rapidly, so that most fitness functions one draws can be compressed to a nearly identical extent). These empirical results show not only the validity of the bound, but that actual behavior has a rapid asymptote; in both the exponential and log-normal cases, the scaling of $N_\mathrm{eff}$ is already sub-logarithmic in $N$.

That the bound relies on the conditional distribution function means that our results are robust to heavy-tailed distributions. What matters is the existence of harmless synonyms; for the states we end up being forbidden to confuse, penalties can be arbitrarily large. We can see this in Fig.~\ref{exp_one}, where we consider a log-normal distribution with mean and variance chosen so that $P(d<0.1\%)$ is identical to the exponential case. The existence of large penalties in the log-normal case does not affect the asymptotic behavior.

This remarkable result has, at first, a counterintuitive feel. As environmental richness rises, it seems that the organism should have to track increasing numbers of states. However, the compression efficiency depends on the difference between environmental uncertainty before and after optimal compression, and these two differences, in the asymptotic limit, scale identically with $N$. 

The existence of asymptotic bounds on memory can also be understood by an example from software engineering. As the web grows in size, a search tool such as Google needs to track an increasing number of pages (environmental complexity rises). If the growth of the web is uniform, and the tool is well-built, however, the number of keywords a user needs to put into the search query does not change over time. For any particular query---``information theory neuroscience'', say---the results returned will vary, as more and more relevant pages are created (ambiguity rises), but the user will be similarly satisfied with the results (error costs remain low).\footnote{We are grateful to Michael Lachmann for suggesting this example.}


\section{The High-Fidelity Regime}

As we approach $d_\mathrm{min}$, the bound given by $1/P(d< D)$ becomes increasingly weak, diverging when $D$ is at the minimal confound level. As we approach this threshold, we enter the high-fidelity regime, where organisms attempt to distinguish difference at a fine-grained level.

In this regime, where we attempt to achieve lower error-rates than those of the minimal confounds, we can no longer allow for strict synonyms between a sub-selection of the off-diagonal elements. Our codebook, $p(r|x)$, must attempt to break the degeneracy between the true environmental state and off-diagonal elements; as the environmental complexity increases, this task, in explicit contrast to the low-fidelity regime, becomes harder as the environment becomes richer.

As in the previous case, we can upper-bound coding costs by construction of a sub-optimal codebook; we can also lower-bound costs by finding the optimal codebook for a strictly less-stringent environment. Coding costs are then bounded between (see Supplementary Information)
\begin{equation}
\left(1-\frac{D}{d_\mathrm{min}}\right)\log_2(N) \leq R(D) \leq \left(1-\frac{D}{\bar{d}}\right)\log_2(N),
\label{scaling}
\end{equation}
where $\bar{d}$ is the average off-diagonal cost in the environment. These bounds hold in the high-fidelity regime where $D\leq d_\mathrm{min}$.

\begin{figure}
\begin{tabular}{cc}
\includegraphics[width=3in]{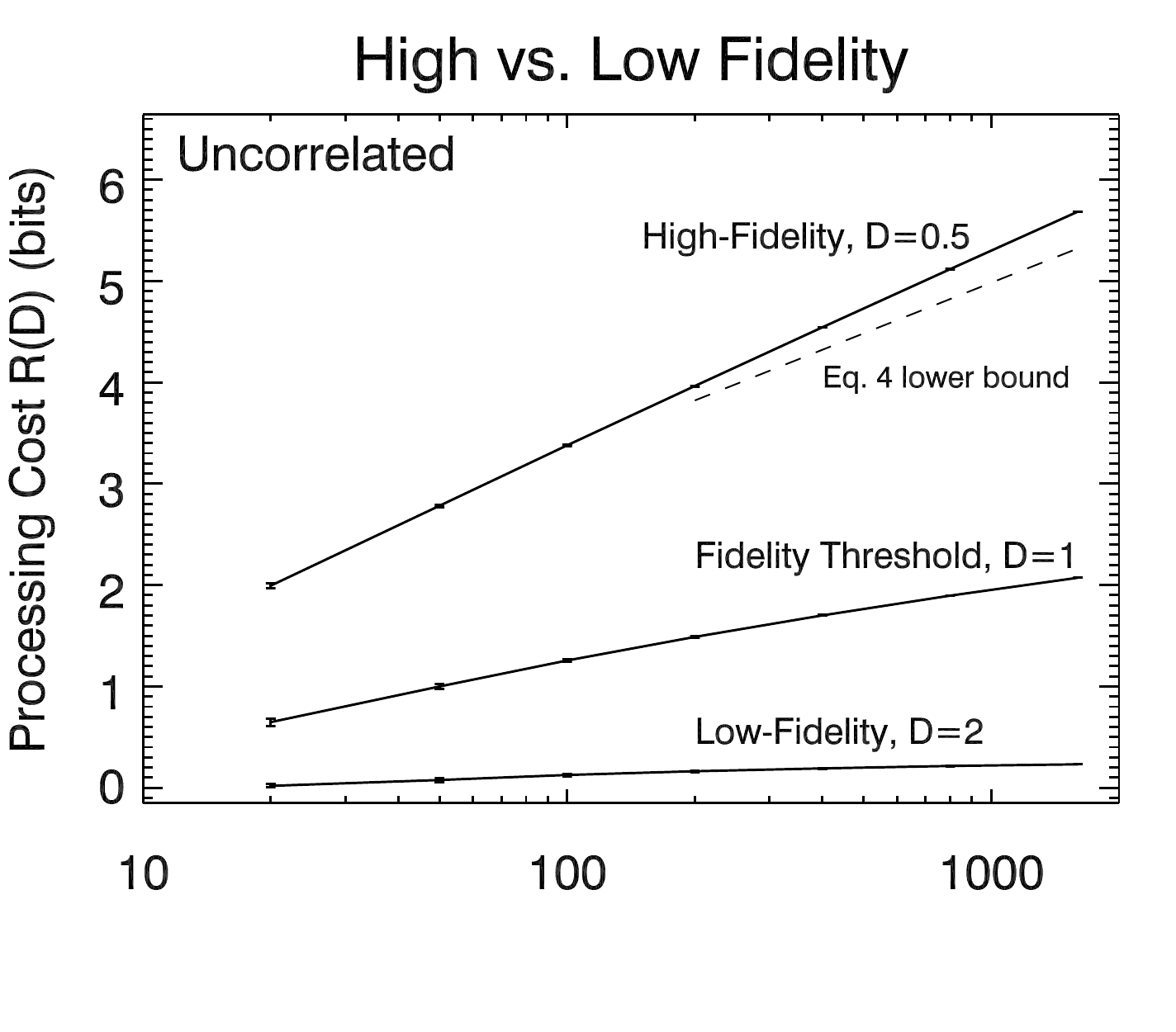} \\
\includegraphics[width=3in]{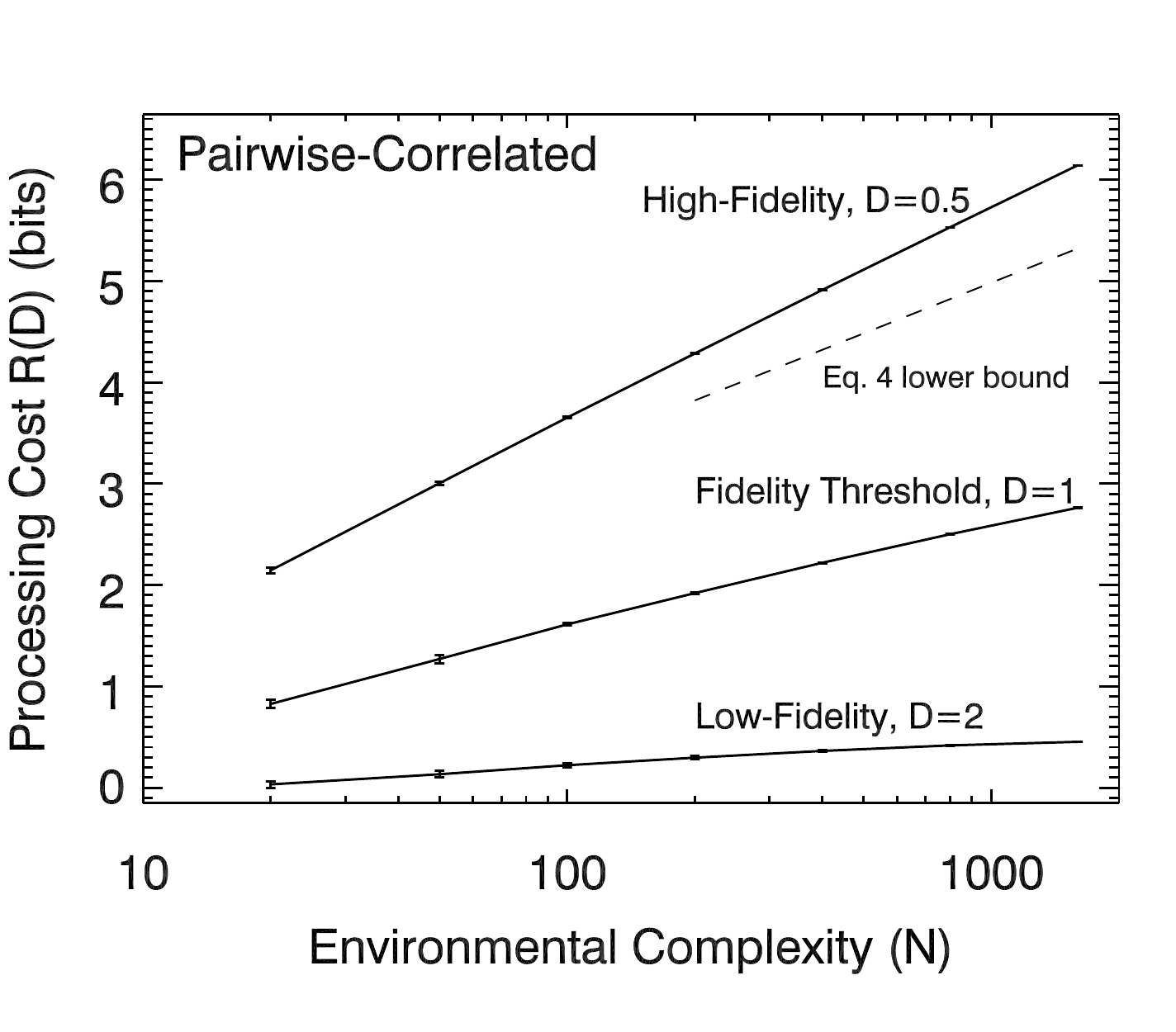}
\end{tabular}
\caption{\textbf{The high-fidelity regime is sensitive to environmental complexity}. Shown here are encoding costs for the shifted exponential distribution, where all errors have a minimal penalty, $d_\mathrm{min}$, of one. When we allow average distortions to be greater than $d_\mathrm{min}$, we are in the low-fidelity regime, and costs, as measured by $R(D)$, become constant. In the high-fidelity regime, when $D$ is less than $d_\mathrm{min}$, costs continue to increase, becoming asymptotically power-law in $N_\mathrm{eff}$, or logarithmic in $R(D)$; the dashed line shows the asymptotic (in this case) square-root scaling for $D/d_\mathrm{min}$ equal to $1/2$. The change-over from constant to this power-law scaling happens around $D\sim d_\mathrm{min}$. A simple argument suggests that evolution under increasing complexity will drive organisms to this transition point.\label{trades}}
\end{figure}

The distributions---exponential and log-normal---described in the previous section do not have a high-fidelity regime; as environmental complexity rises, so does the number of harmless synonyms. If, however, we put a floor on these distributions, so that there is always a strictly non-zero penalty for confusion, the scaling of Eq.~\ref{scaling} appears. Fig.~\ref{trades} compares these two regimes---the low-fidelity regime of Eq.~\ref{constant}, and the high-fidelity regime of Eq.~\ref{scaling}---for error costs with $d(x,\tilde{x})$ greater than unity, and distributed exponentially. In the low-fidelity regime, $D > 1$, we see the same asymptotic complexity as in Fig.~\ref{exp_one}; when we try to achieve bounds below the minimal confound, the lower bound of Eq.~\ref{scaling} becomes relevant, and $R(D)$ scales logarithmically with $N$ (equivalently, $N_{\mathrm{eff}}$ is a power-law in $N$, with $N_{\mathrm{eff}}$ equal to $N$ in the zero error limit). 

The change-over is shown graphically in Fig.~\ref{trades}. While processing costs for the low-fidelity regime asymptote to a constant as $N$ becomes large, the lower bound for the high-fidelity regime provides an asymptotic limit to the slope of the $N_\mathrm{eff}$ curve on a log-log plot (equivalently, a $R(D)\sim\log{N}$ relationship). This holds true even when environmental costs are correlated; the lower panel of Fig.~\ref{trades} shows the same curves where the unconditioned cost distributions are identical, but we have induced pairwise correlations by averaging each row with a shuffled version of itself (except for the diagonal point, which remains zero).

\section{Evolution and Poised Perception}

An organism's fitness includes both its error rate, $D$, and the processing costs associated with achieving that rate, $R(D)$. Under the simplifying assumption that these costs add independently, and taking processing costs to be proportional to $R(D)$, we can take an organism's fitness, $F$ to be a monotonically decreasing function of total cost $D+\beta^{-1} R(D)$, which takes into account the costs of both error and perceptual processing. The Lagrange multiplier, $\beta$, characterises the error--computation tradeoff. This basic linear tradeoff is a natural choice that has been the basis of studies of encoding in molecular biological systems (Ref.~\cite{tlusty2008rate} and references therein). 

Evolutionary processes tend to maximize fitness at the population level; the fitness of an organism can be thought of as its reproductive rate relative to average. Whatever the value of $\beta$, a higher than average fitness will, given sufficiently faithful reproduction, translate into exponential gains for that organism and its descendants. Rather than considering absolute differences we can then focus on rank order---what changes cause the fitness to rise or fall relative to baseline.

For any particular $\beta$ tradeoff, there will be a unique, fitness-optimal solution. Because fitness is convex in the choice of the codebook~\cite[Thm.~3]{Tish00a}, finding the optimal point corresponds to an unsupervised classification of sensory signals based on gradient descent. It is thus a natural task for an organism's neural system~\cite{hinton1989connectionist,back1993overview,amari1998natural,Friston2009293}, and also corresponds to the simplest models of selection in evolutionary time~\cite{Frank2012-FRAWAL}. 

By finding new and more efficient mechanisms to process information, evolution can increase an individual's $\beta$.\footnote{We assume the timescale on which $\beta$ changes is much slower than the timescale for an organism to find the optimal codebook at fixed $\beta$. This is natural when, for example, codebooks are found in ontogenetic time while improvements in the information processing substrate must evolve.} This enables the individual to make finer and finer distinctions and thereby both lower $D$ and raise its fitness. In the low-fidelity regime, when organisms compete in this fashion, their gains in accuracy are independent of increases in environmental complexity. As they increase $\beta$ through evolutionary innovation, individuals will find an optimal fitness point at lower and lower values of $D$. Above the $d_\mathrm{min}$ threshold, even large changes in $N$ will leave the optimal $D$ for any particular $\beta$ unchanged. Low-fidelity perceptual systems are robust to large changes in environmental complexity.


As $D$ approaches, and then goes below, $d_\mathrm{min}$, the picture changes drastically. Organisms may compete to lower $D$, but gains are transitory when the environment is itself changing. In particular, increases in environmental complexity mean that $R(D)$ will now rise at least logarithmically as $N$ increases. For the lower bound, we have
\begin{equation}
\frac{\partial^2}{\partial N\partial D}(D+\beta^{-1}R(D))\leq -\frac{1}{d_\mathrm{min}\beta N} <0,
\end{equation}
so that, when $N$ increases, organisms at the lower bound can lower costs by going to larger $D$. (This derivative relationship holds even when the bound is not tight, but $R(D)$ takes the form $f(D)\log_2{N}$, with $f(D)$ independent of $N$, since $R(D)$ is a decreasing function of $D$. More generally still, as $N$ increases, the growing lower bound of Eq.~\ref{scaling} means that costs in the high-fidelity regime will eventually exceed the fixed costs of solutions around $D\sim d_\mathrm{min}$.)


Since fitness is a monotonic function of cost, the new fitness maximum will also lie in the direction of larger $D$. As complexity rises, organisms in the high-fidelity regime will be driven towards $d_\mathrm{min}$, at the threshold between high and low-fidelity perception. This driving force, towards larger $D$, also holds when we take the costs to scale with effective symbol number, $N_\mathrm{eff}$, which is exponential in $R(D)$.

Taken together, these results imply that, when organisms compete in an environment of rising complexity, we expect to find their perceptual apparatus poised around $D\sim d_\mathrm{min}$. We refer to this as poised perception.

Organisms will compete on processing efficiency, increasing $\beta$ and pushing $D$ downward. This process is asymptotically insensitive to environmental complexity until a transition region around $d_\mathrm{min}$. Beyond this point, rising complexity provides a counter-force. When $N$ is large and rising on timescales faster than those on which evolution increases $\beta$, we expect organisms to stall out at this threshold. Even when they are poised around $d_\mathrm{min}$, increasing $\beta$ will still give an individual an evolutionary advantage; the effect of a changing environment is to force organisms to compete on efficiency, rather than richness of internal representations.

\section{Discussion}

Where there is an accuracy--memory tradeoff, our results suggest that organisms in environments of rising complexity will find themselves able to distinguish minimal confounds, but no more. 

We expect this to happen when the number of categories to be distinguished can rise. It is worth considering what happens when the environment has a hierarchical structure, and where there may be a fixed number of coarse-grained distinctions that are crucial to maintain, such as those between predator and prey, kin and non-kin, in-group and out-group.

Because of this hierarchical structure we do not expect organisms to have a single system for representing their environment. The results here then apply to subdomains of the perceptual problem. In the case of predator--prey, for example, we expect optimal codebooks will look block-diagonal, with a clear predator--prey distinction, and then subdistinctions for the two categories (eagle vs.~leopard; inedible vs.~edible plants). 

For the large distinction, $N$ is fixed, but environmental complexity can increase within each category. As new forms of predators or prey arise, the arguments here suggest that organisms will be poised at the thresholds within each subsystem. When, for example, prey may be toxic, predators will evolve to distinguish toxic from non-toxic prey, but tend not to distinguish between near-synonyms {\it within} either the toxic or non-toxic categories. 

Such an effect may influence the evolution of Batesian mimicry---where a non-toxic prey species imitates a toxic one---when such mimicry is driven by the perceptual abilities of predators~\cite{2013review}. Our arguments here suggest that predator perception evolves in such a way that Batesian mimics will have a phenotypic range similar to that found between two toxic species. A harmless Batesian mimic can then emerge if it can approximate, in appearance, a toxic species by roughly the same amount as that species resembles a second species, also toxic. This also suggests that a diversity of equally-toxic species will lead to less-accurate mimicry. This is not because a mimic will attempt to model different species simultaneously (the multimodal hypothesis~\cite{edmunds2000there}), rather that predators who attempt to make finer-grained distinctions within the toxic-species space will find themselves driven back to $d_\mathrm{min}$ when the total number of species increases.

Our results also have implications for the cognitive and social sciences. A small number of coarse-grained distinctions are often found in human social cognition, where we expect a block-diagonal structure over a small, fixed number of categories (kin~vs.~non-kin, in-group~vs.~out-group). Within each of the large distinctions, our representational systems are then tasked with making fine-grained distinctions over sets of varying size. 

When we attempt to make distinctions within a sub-category of increasing size, our results suggest that we will be barely competent at making the least important distinctions. Informally, we are just barely able to distinguish our least-important friends when our circles expand. Such ideas might be tested in a laboratory-based study that trains subjects on a set of increasingly difficult distinction tasks with varying rewards. The results here predict that, as they improve their decision-making abilities, individuals will develop representations that are barely able to make distinctions between the least-important cases. 

Our results on poised perception apply when environments can increase in complexity over time and when perceptual systems struggle to represent and process environmental states. Rate-distortion theory provides new quantitative insight into the underlying structure of this problem. While not all perception problems take this form, those that do are key evolutionary drivers of increased perceptual and cognitive ability.

\vspace{0.5cm}
\begin{acknowledgments}
This research was supported by National Science Foundation Grant \#EF-1137929. We thank John M.~Beggs,  Jonathon D.~Crystal, Michael Lachmann, Chris Hillar, Jim Crutchfield, Susanne Still and Mike DeWeese for helpful discussion. S.M. acknowledges the support of a University of California Berkeley Chancellor's Fellowship, and of an NSF Graduate Research Fellowship.

\end{acknowledgments}

\onecolumngrid
\appendix
\clearpage
\section{Supplementary Information}

In this Appendix, we provide more discussion of the basic rate-distortion set-up and connect it to recent progress in understanding the sensorimotor loop. We present extended derivations of the results presented in the main text. And we present results that suggest, in the low-fidelity regime, a distribution-dependent universality in the rate-distortion curve

We begin with a typical reinforcement learning setup~\cite{barto1998reinforcement}.  An agent takes a total of $T$ actions $a_0,\ldots,a_T$ throughout its life in response to an environment whose state at any given timestep $i$ is $x_i$.  The actions that the agent takes are, we assume, dependent only on the sensory percepts that the agent has stored at timestep $i$, $\tilde{x}_i$.  We denote this by writing $a_i$ as $a_i(\tilde{x}_i)$.  In principle, these percepts can take into account information about the environment's past as well as its current state.  The reward that an agent receives upon taking a particular action $a_i(\tilde{x}_i)$ is taken to be a function of the environment's present state $x_i$, $r(x_i,a_i(\tilde{x}_i))$.  The agent accumulates a total reward of
\begin{equation}
R = \sum_{i=0}^T r(x_i,a_i(\tilde{x}_i))
\end{equation}
over its lifetime. In principle, this reward function could depend upon the environment's past as well as its present state \cite{Stil07c, tishby2011information, singh2004predictive, Brodu11}; this additional layer of complexity is unnecessary to make the present point.

Much work in reinforcement learning has focused on choosing optimal action policies. However, good action policies require accurate or nearly-accurate sensory perceptions.  If we mistake a lion for a deer, we will probably die, despite having an optimal plan in place once we see a deer. We thus limits to our sensory models given finite resource constraints. 

When the joint time series of inputs and actions is stationary, we can move from viewing the environment and the agent's actions as time series to viewing them as realizations of a stochastic process characterized by probability distributions $\Prob(X=x,\tilde{X}=\tilde{x},A=a)$.  This probability distribution implicitly allows for random action policies; a deterministic action policy would imply that $\Prob(A=a|\tilde{X}=\tilde{x}) = \delta_{a,a(\tilde{x})}$.  The assumption from earlier that the action policy depends on the environment only through the sensory percept is more compactly described as $X\rightarrow\tilde{X}\rightarrow A$ when we view the environmental state, the agent's sensory percept, and the agent's actions as random variables.  And finally, the total reward is now expressible as
\begin{equation}
R \approx T \sum_{x,\tilde{x},a} \Prob(X=x,\tilde{X}=\tilde{x},A=a) r(x,a(\tilde{x})),
\end{equation}
so we find it convenient to define an expected reward of $\mathbb{E}[r] = \frac{R}{T} \approx \sum_{x,\tilde{x},a} \Prob(X=x,\tilde{X}=\tilde{x},A=a) r(x,a(\tilde{x}))$.

Many real-world situations ask us to model environments which are prohibitively large, and recording the present environmental state with perfect fidelity is unreasonable given finite memory, energy, and other resource constraints.  These constraints are thought to be relevant to how organisms actually function, \emph{e.g.}, as in Refs. \cite{attwell2001energy, chklovskii2004maps, raichle2002appraising, hasenstaub2010metabolic}, though the particular way in which these resource constraints affect organism computation might depend strongly on the organism's specific memory substrate.  For example, \textit{E. coli} bacteria, which store information about concentration gradients in the stochastic behavior of their ligand-gated ion channels, will have different energy, timing, and space costs than a human who stores information about more complex environments in everything from the structure of dendritic trees to the higher-level structure of cortical columns and all the way up to the maintenance of external records made possible by symbolic systems and writing.

Despite this, rate-distortion theory provides insight into the tradeoff between the quality of an agent's perception of the environment and the memory required to store those perceptions, regardless of the particular biological substrate for that memory.  In economics, this realization goes under the name of ``rational inattention'' \cite{sims2003implications}; in theoretical neuroscience, this idea has taken hold in a few recent papers \cite{Palm13a, sims2015cost}, though more attention has been focused on the lossless limit \cite{simoncelli2001natural}.  These ideas are related to, but distinct from, Bayesian inference.  We focus only on the memory limitations related to storing sensory information and not those related to taking action; see Refs.~\cite{Stil07c, tishby2011information} and references therein for efforts on the latter.

To make these general ideas concrete, we focus on an agent with $n$ neurons which communicate sensory information to higher cortical levels by their collective state at each time step.  There are a total of $2^n$ possible states of these $n$ neurons, implying a maximal rate of $\log_2 2^n = n$ bits per input symbol.  For more direct connection with the language of rate-distortion theory, we think about our actions not as opportunities to reap rewards, but as potentially lost opportunities to reap higher rewards.  If we fail to faithfully record an environmental state $x$ with the sensory percept $\tilde{x}$, then our actions will necessarily fall short of the optimal actions.  For each environmental state $x$ and action policy $\Prob(A|\tilde{X})$, there is some maximum reward $r^*(x)$ achieved by faithfully recording $x$ in our sensory percept with an expected maximal reward of
\begin{equation}
\mathbb{E}[r^*] = \sum_{x} \Prob(X=x) r^*(x).
\end{equation}
The rate-distortion theorem implies that
\begin{equation}
\min_{\Prob(\tilde{X}|X):\mathbb{E}[r^*]-\mathbb{E}[r]\leq D} I[\mathcal{R};X] = R(D) \leq n,
\end{equation}
or, in words, that the implicitly action policy dependent rate-distortion function is a lower bound on the number of sensory neurons.
For details, see Refs. \cite{Yeun08a} or other standard information theory texts.  This bound may or may not be tight, though there is preliminary evidence that the bound is surprisingly tight \cite{Palm13a, sims2015cost}.  We emphasize that this example was used to illustrate an even more general point.  In other setups, this rate-distortion function may be a lower bound on the number of hidden states in an ion channel, for instance.

The questions that we ask in this manuscript do not require that this bound be tight; rather, we aim to ask how $R(D)$ scales with the number of environmental states $N$, and hope that the scaling relations provided are tight even if the specific bound is not.  A key quantity, $d_\mathrm{min}$, is the minimal price we pay for not faithfully transmitting the environmental state:
\begin{equation}
d_\mathrm{min} := \min_{x,\tilde{x}\neq x} r^*(x)-r(x,a(\tilde{x})),
\end{equation}
and pairs that satisfy this bound are called ``minimal confounds''. Despite the generality of this setup, there are apparently two different regimes for how $R(D)$ scales with $N$.  When $D\leq d_\mathrm{min}$, then $R(D)$ scales logarithmically with $N$; when $D>d_\mathrm{min}$, then $R(D)$ asymptotes to a constant as $N$ tends to infinity.  (This somewhat surprising result might be anticipated by an earlier theorem \cite{Rose94a} showing that the optimal lossy representations of continuous random variables are discrete.)  Additional structure in the environment would likely contribute to delineation of further scaling regimes, which we do not pursue here.

If we assume that we should choose the minimal $D$ such that we have some robustness to sudden increases in environmental complexity---\emph{i.e.}, such that $R(D)$ does not increase greatly with $N$---then our results suggest that $D$ is optimally set to be $d_\mathrm{min}$.
\\

\subsection{An example: random environments}

To obtain some intuition, we consider the case that distortions are chosen i.i.d.~from a distribution with probability density function $\phi$.  (From our earlier construction, on-diagonal entries are set to $d(x,x) = 0$.)  

See Fig. $1$ in the main text, giving $R(D)$ for several $D$ for various distributions $\phi$ and $N$. These numerical experiments suggest a weak universality in rate-distortion functions, apparent from the increasingly small variance in $R(D)$ for different random distortion matrices: as $N\rightarrow\infty$, $R(D)$ is a statistic of $\phi$ and is independent of the specific distortion measure.  If true, this conjecture implies that the required resources might be time-invariant even if the reward function is not.  




\begin{figure*}
\includegraphics[width=0.45\textwidth]{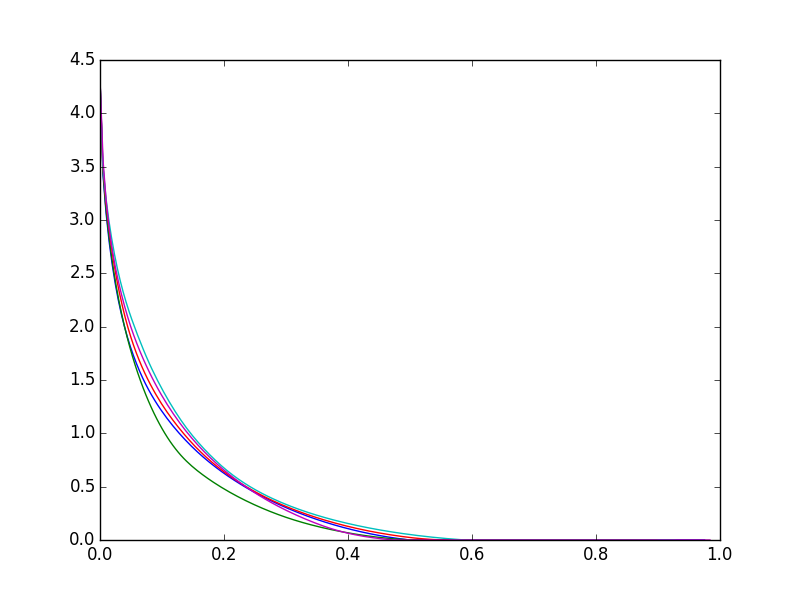}
\includegraphics[width=0.45\textwidth]{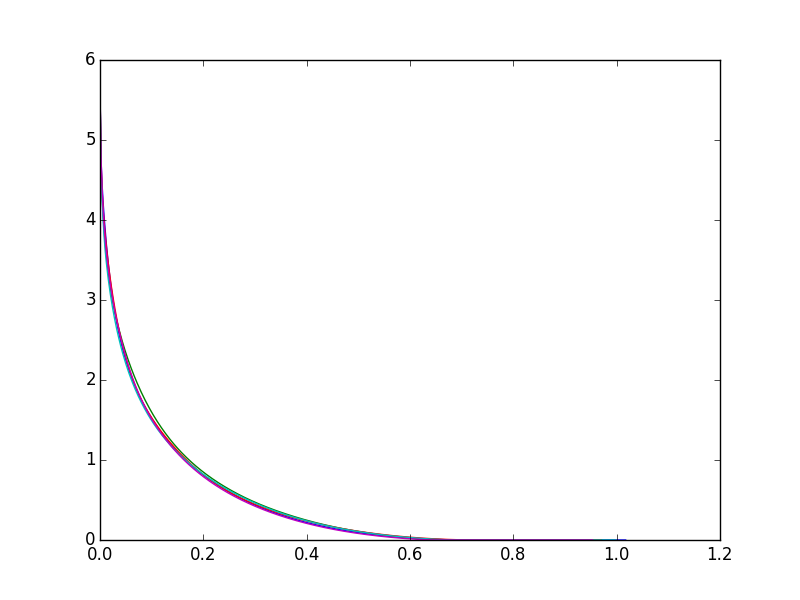}
\includegraphics[width=0.45\textwidth]{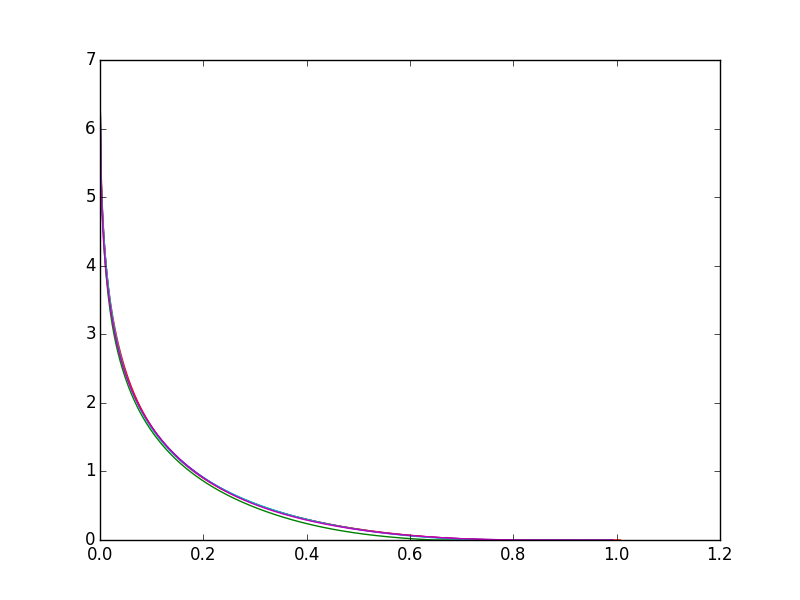}
\includegraphics[width=0.45\textwidth]{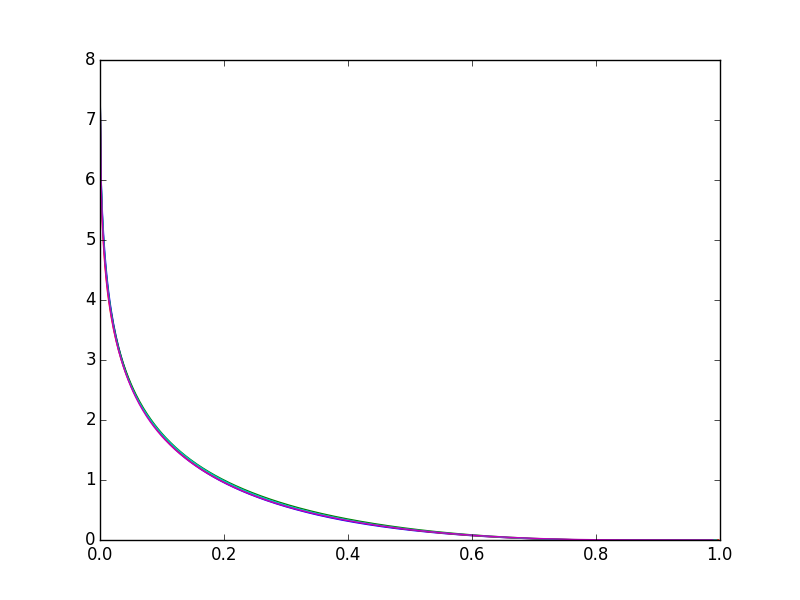}
\includegraphics[width=0.45\textwidth]{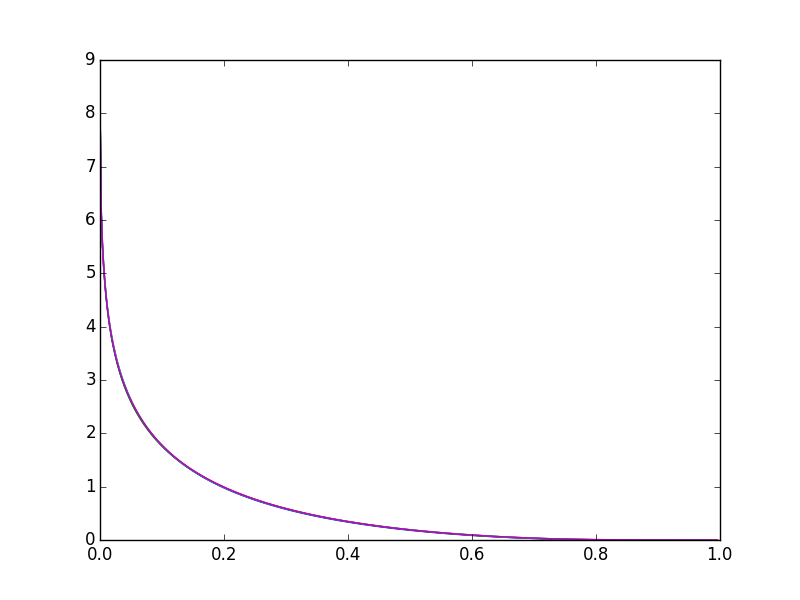}
\includegraphics[width=0.45\textwidth]{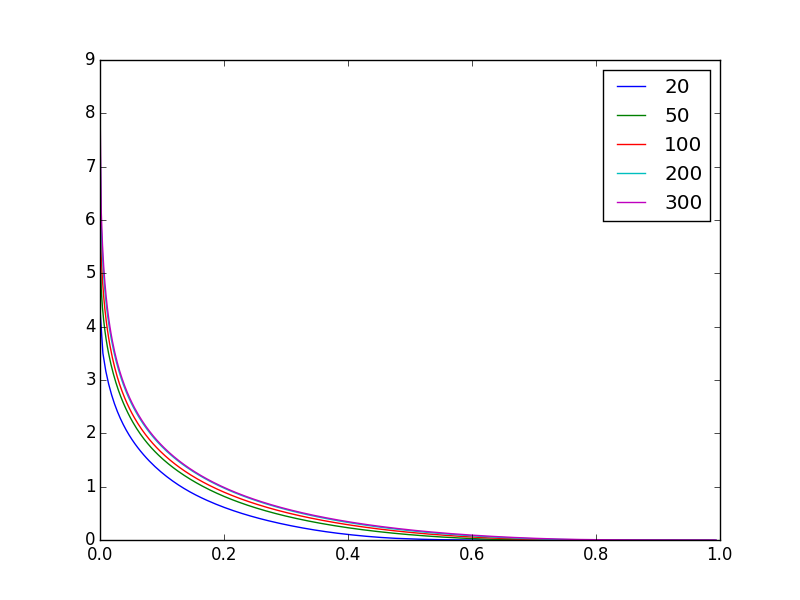}
\caption{\textbf{Rate-distortion functions for $\phi(x)= e^{-x}$ for increasing $N$.} {\bf First five plots}: rate-distortion functions for five different samples of the distortion measure; from left to right, we show $N$ of 20, 50, 100, 200 and 300. Rate-distortion functions for different draws from the $d(x,\tilde{x})$ ensemble are nearly identical as $N$ gets large.  {\bf Bottom right plot}: the means of the rate-distortion functions over the ensemble of matrices for various $N$.  As $N$ increases, these appear to limit to a rate-distortion function which is everywhere bounded.}
\label{fig:Exp11}
\end{figure*}


\begin{figure*}
\includegraphics[width=0.45\textwidth]{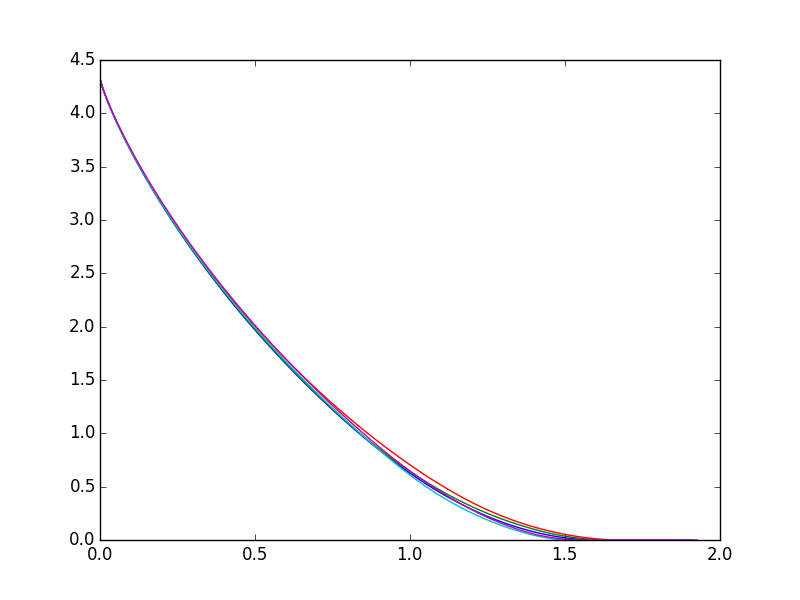}
\includegraphics[width=0.45\textwidth]{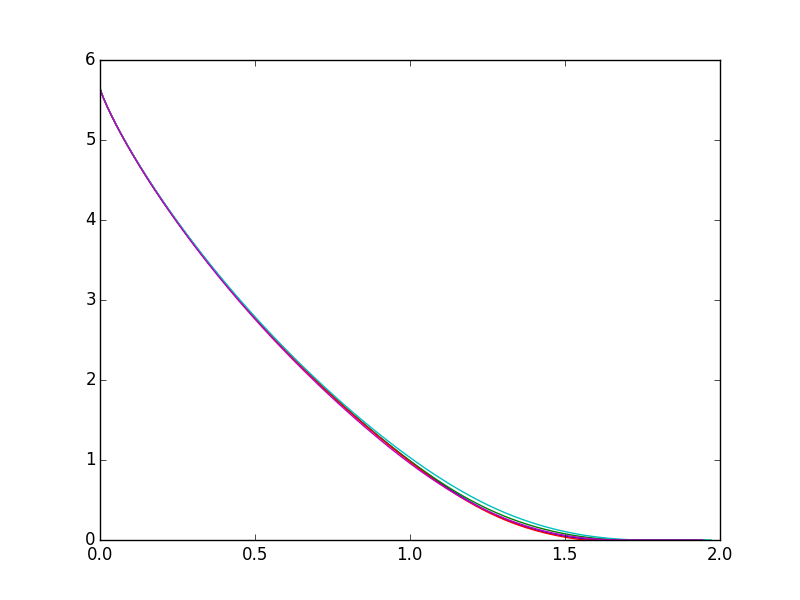}
\includegraphics[width=0.45\textwidth]{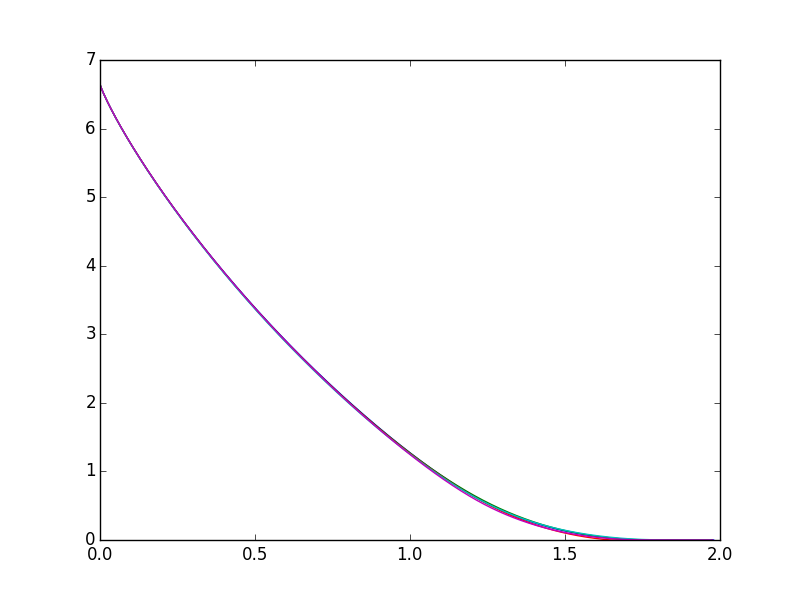}
\includegraphics[width=0.45\textwidth]{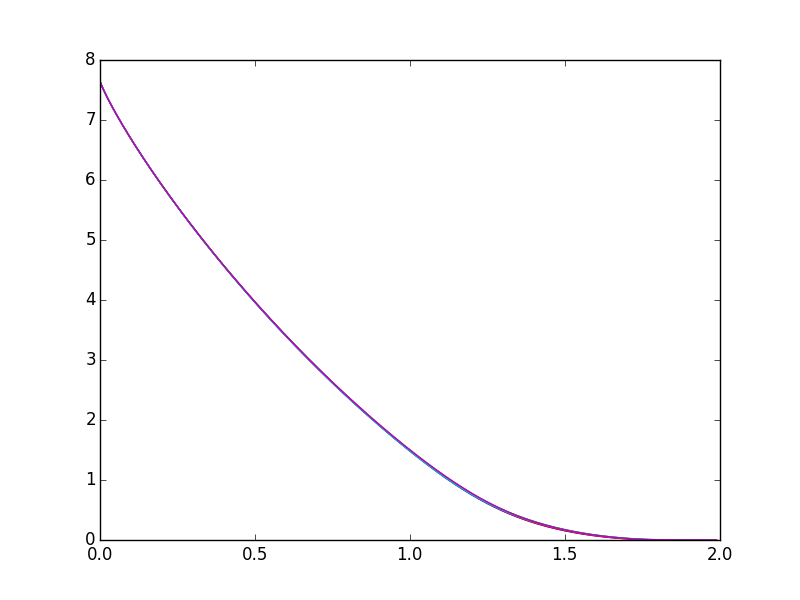}
\includegraphics[width=0.45\textwidth]{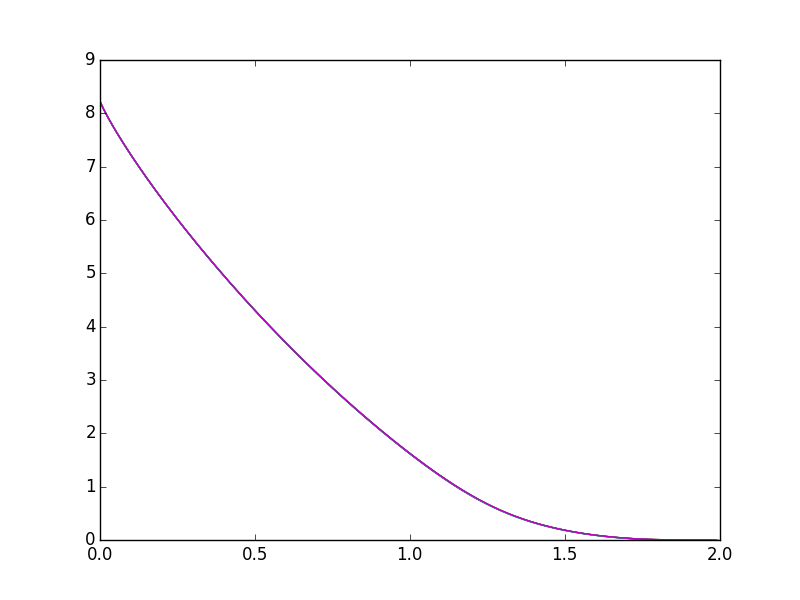}
\includegraphics[width=0.45\textwidth]{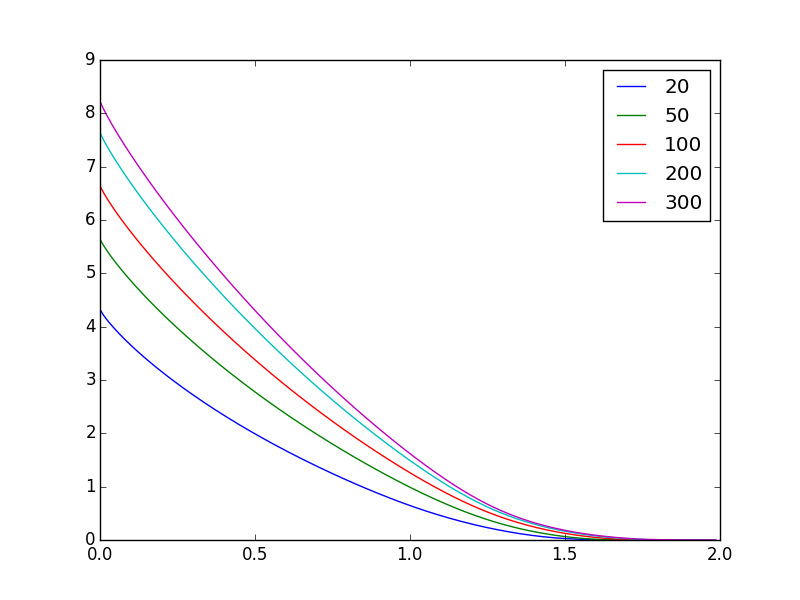}
\caption{\textbf{Rate-distortion functions for $\phi(x)= 1+e^{-x}$ for increasing $N$.}  {\bf First five plots}: rate-distortion functions for five different samples of the distortion measure; from left to right, we show $N$ of 20, 50, 100, 200 and 300.  At $N\geq 100$, rate-distortion functions for different draws from the ensemble are nearly identical.  {\bf Bottom right plot}: the means of the rate-distortion functions over the ensemble of matrices for various $N$.  As $N$ increases, the behavior is distinct from that in Fig.~\ref{fig:Exp11} and $R(D)$ curves for increasing $N$ do not asymptote to a single distribution; for $D\leq 1$, analytic bounds find that $R(D)$ will increase without bound as $N$ increases; spacing between lines should scale logarithmically (Eq.~4, main text).}
\label{fig:Exp21}
\end{figure*}

We now ask how $\phi$ affects the scaling of $R(D)$ with $N$. When we mean-shift $\phi$ by $d_\mathrm{min}$, $R(D)$ starts to show deviations from the weak universality just described.  Two different random distortion measures with the same $\phi$ and $N$ still produce nearly identical $R(D)$.  But for expected distortions $D< d_\mathrm{min}$, $R(D)$ appears to grow logarithmically with $N$, while for expected distortions $D\geq d_\mathrm{min}$, $R(D)$ tends to a constant, independent of $N$. This is shown in Fig.~\ref{fig:Exp21}.

\subsection{Scaling of $n$ with $N$ depends on desired reward}

We can use analytic upper and lower bounds to explain some of the results found above.  Define the minimal distortion
\begin{equation}
d_\mathrm{min} := \min_{x\neq\tilde{x}} d(x,\tilde{x})
\end{equation}
and the average distortion
\begin{equation}
\bar{d} := \frac{1}{N(N-1)} \sum_{x\neq\tilde{x}} d(x,\tilde{x}).
\end{equation}
The former is the distortion in the best-possible miscoding.  The latter is the expected distortion when we choose our actions independently of the environment.  Also, assume that inputs are equiprobable, $p(x) = \frac{1}{N}$ for all $x$.

First, we upper bound the rate-distortion function in the high-fidelity regime using the suboptimal codebook
\begin{equation}
p(\tilde{x}|x) = \begin{cases} 1-C & \tilde{x}=x \\ \frac{C}{N-1} & \tilde{x}\neq x \end{cases}
\end{equation}
for $0\leq C\leq 1$.  Symmetry implies that $p(\tilde{x}) = \frac{1}{N}$ for all $\tilde{x}$.  This codebook has an expected distortion of
\begin{equation}
\langle d\rangle = \sum_{x,\tilde{x}} p(x) p(\tilde{x}|x) d(x,\tilde{x}) = \frac{C}{N(N-1)} \sum_{x\neq\tilde{x}} d(x,\tilde{x}) = C\bar{d},
\end{equation}
and a rate of
\begin{equation}
I[X;\tilde{X}] = H[\tilde{X}]-H[\tilde{X}|X] = \log_2 N - \frac{1}{N} \sum_x H[\tilde{X}|X=x]
\end{equation}
where application of the additivity axiom shows
\begin{equation}
H[\tilde{X}|X=x] = H_b[C] + C \log_2 (N-1).
\end{equation}
Together, this gives
\begin{equation}
I[X;\tilde{X}] = \log_2 N - H_b[C] - C \log_2 (N-1) = \log_2 N - H_b\left[\frac{\langle d\rangle}{\bar{d}}\right] - \frac{\langle d\rangle}{\bar{d}}\log_2 (N-1).
\end{equation}
Hence, when $D\leq \bar{d}$, the rate-distortion function is upper-bounded by
\begin{equation}
R(D) \leq \log_2 N - H_b\left[\frac{D}{\bar{d}}\right] - \frac{D}{\bar{d}} \log_2 (N-1) \approx \left(1-\frac{D}{\bar{d}}\right)\log_2 N.
\end{equation}
Next, we lower-bound the rate-distortion function by lower-bounding the distortion-rate function, 
\begin{equation}
D(R) := \min_{p(\tilde{x}|x):I[X;\tilde{X}]\leq R} \langle d\rangle.
\end{equation}
This is (by construction) greater than or equal to the distortion-rate function for the distortion measure,
\begin{equation}
d_L(x,\tilde{x}) = \begin{cases} 0 & x=\tilde{x} \\ d_\mathrm{min} & x\neq \tilde{x} \end{cases}.
\end{equation}
By symmetry, the optimal codebook for this distortion measure takes the form of the suboptimal codebooks above.  (It is straightforward to check that those suboptimal codebooks satisfy the nonlinear equations in Thm. $3$ of Ref. \cite{Tish00a}.)  Hence, the rate-distortion function is lower-bounded:
\begin{equation}
\log_2 N - H_b\left[\frac{D}{d_\mathrm{min}}\right] - \frac{D}{d_\mathrm{min}} \log_2 (N-1) \leq R(D),
\end{equation}
valid when $D\leq d_\mathrm{min}$.
Together, the two bounds imply
\begin{equation}
\log_2 N - H_b\left[\frac{D}{d_\mathrm{min}}\right] - \frac{D}{d_\mathrm{min}} \log_2 (N-1) \leq R(D) \leq \log_2 N - H_b\left[\frac{D}{\bar{d}}\right] - \frac{D}{\bar{d}} \log_2 (N-1)
\end{equation}
when $D\leq d_\mathrm{min}$.
In the large $N$ limit, we find that
\begin{equation}
\left(1-\frac{D}{d_\mathrm{min}}\right) \log_2 N \leq R(D) \leq \left(1-\frac{D}{\bar{d}}\right) \log_2 N.
\end{equation}
So, the rate-distortion function and the required number of neurons scale logarithmically with the number of inputs in the high-fidelity regime.
%
\noindent \subsection{Low-fidelity regime}

Now we find an upper bound on the rate-distortion function in the regime that $D\geq d_\mathrm{min}$.  Consider a suboptimal codebook of the form
\begin{equation}
p(\tilde{x}|x) = \begin{cases} \frac{1}{N_D(x)} & d(x,\tilde{x})\leq D \\ 0 & d(x,\tilde{x})>D \end{cases}
\end{equation}
where
\begin{equation}
N_D(x) := \sum_{\tilde{x}} \delta_{d(x,\tilde{x}) \leq D}.
\end{equation}
By construction, this codebook has an expected distortion $\leq D$.  Its rate is
\begin{equation}
I[X;\tilde{X}] = H[\tilde{X}] - H[\tilde{X}|X] = H[\tilde{X}] - \frac{1}{N} \sum_x \log_2 N_D(x).
\end{equation}
In general, $p(\tilde{x})$ could be rather complicated, but we can always upper bound $H[\tilde{X}]\leq\log_2 N$, giving
\begin{equation}
I[X;\tilde{X}] \leq \log_2 N - \frac{1}{N} \sum_x \log_2 N_D(x) = \frac{1}{N} \sum_x \log_2 \frac{N}{N_D(x)}.
\end{equation}
As long as $N_D(x)$ scales linearly with $N$ and not sublinearly, then the rate of this codebook is independent of $N$ in the large $N$ limit.  Hence we find an upper bound
\begin{equation}
R(D) \leq \frac{1}{N} \sum_x \log_2 \frac{N}{N_D(x)}.
\end{equation}
For many distortion measures---those for which $N_D(x) \propto N$---this upper bound will imply that $R(D)$ is $O(1)$ in $N$.  For instance, if distortions are chosen randomly (as we discussed earlier), then
\begin{equation}
\lim_{N\rightarrow\infty}\frac{N}{N_D(x)} = \frac{1}{\Prob(d(x,\tilde{x})\leq D)}
\end{equation}
and
\begin{equation}
R(D) \leq \log_2 \frac{1}{\Prob(d(x,\tilde{x})\leq D)}.
\end{equation}
The upper bound relies only on a single feature of the probability distribution, not on $N$.

\clearpage

\section*{References}
\bibliography{arxiv_version}

\end{document}